\documentclass[aps,pra,showpacs]{revtex4}
\usepackage{graphicx,color}
\usepackage{hyperref}

\begin{document}

\title{Correlation functions of a Lieb-Liniger Bose gas}

\author{G.E. Astrakharchik$^{(1,2)}$ and S. Giorgini$^{(1)}$}

\affiliation{
$^{1}$Dipartimento di Fisica, Universit\`a di Trento, and BEC-INFM, I-38050 Povo, Italy\\
$^{(2)}$Institute of Spectroscopy, 142190 Troitsk, Moscow region, Russia}

\date{\today}

\begin{abstract}
The ground-state correlation functions of a one-dimensional homogeneous Bose system
described by the Lieb-Liniger Hamiltonian are investigated by using exact quantum
Monte Carlo techniques. This article is an extension of a previous study published
in Phys. Rev. A {\bf 68}, 031602 (2003). New results on the local three-body correlator 
as a function of the interaction strength are included and compared with the measured 
value from three-body loss experiments. We also carry out a thorough study of the 
short- and long-range behavior of the one-body density matrix.
\end{abstract}
\pacs{03.75.Fi, 05.30.Fk, 67.40.Db}

\maketitle

\section{Introduction}

Recent progress achieved in techniques of confining Bose condensates has lead to
experimental realizations of quasi-one-dimensional (1D)
systems\cite{MIT, Paris, Zurich, Laburthe Tolra, Paredes, PennStateU}. The quasi-1D
regime is reached in highly anisotropic traps, where the axial motion of the atoms
is weakly confined while the radial motion is frozen to zero point oscillations by
the tight transverse trapping. These experimental
achievements have revived interest in the theoretical study of the properties of 1D 
Bose gases. In most applications, a single parameter, the
effective 1D scattering length $a_{1D}$, is sufficient to describe the
interatomic potential, which in this case can be conveniently modeled by a
$\delta$-function pseudopotential. For repulsive effective interactions the relevant
model is provided by the Lieb-Liniger Hamiltonian \cite{LL1}. Many properties of
this integrable model such as the ground-state energy \cite{LL1}, the excitation
spectrum \cite{LL2} and the thermodynamic functions at finite temperature \cite{YY} were
obtained exactly in the 60' using the Bethe {\it ansatz} method. Less is known about
correlation functions, for which analytic results were obtained only in the strongly
interacting regime of impenetrable bosons \cite{Corr tonks} and for the long-range behavior
of the one-body density matrix~\cite{Schwartz}. More recently, the
properties of correlation functions of the Lieb-Liniger model have attracted
considerable attention and the short-range expansion of the one-body density matrix
\cite{Olshanii}, as well as the local two- and three-body correlation function
\cite{Gangardt} have been investigated. However, a precise determination of the
spatial variation of correlation functions for arbitrary interaction strength is
lacking.

We use exact quantum Monte Carlo methods to investigate the behavior of correlation 
functions in the ground state of the Lieb-Liniger model. Over a wide range of values 
for the interaction strength, we calculate the one- and two-body correlation function
and their Fourier transform giving, respectively, the momentum distribution and the static 
structure factor of the system. These results were already presented in a previous study 
(Ref.~\cite{us}) and are briefly reviewed here. We present new results on the local 
three-body correlation function ranging from the weakly- up to the strongly-interacting 
regime. We have also investigated in more details the long- and short-range behavior of 
the one-body density matrix, including a discussion of finite-size effects and of the 
validity of the trial function used for importance sampling. We always provide 
quantitative comparisons with known analytical results obtained in the weak- or 
strong-correlation regime, or holding at short or large distances. In the case of the
local three-body correlator we also compare with available experimental results 
obtained from three-body loss measurements~\cite{Laburthe Tolra}.     

The structure of the paper is as follows. In section \ref{FF} we introduce the
definitions of the spatial correlation functions and their Fourier transform. 
In section \ref{LL} we discuss the Lieb-Liniger model and give a summary of the 
main known results concerning correlation functions in this model.
Section \ref{SMC} is devoted to a brief description of the quantum Monte Carlo
method used for the numerical solution of the Schr\"odinger equation. The
optimization of the trial wave function used for importance sampling and the effects
due to finite-size are discussed here. The results for correlation functions are 
presented in section \ref{results}. Finally, in section \ref{SC} we draw our conclusions.

\section{Correlation functions \label{FF}}

We use the first quantization definition of correlation functions in terms of the
many-body wave function of the system $\Psi(z_1, ..., z_N)$, where $z_1, ..., z_N$
denote the coordinates of the $N$ particles. We are interested in the regime where
the most relevant fluctuations are of quantum nature. In the following we will
always consider homogeneous systems at $T=0$ and the ground-state wave function of
the system will be denoted by $\Psi_0(z_1, ..., z_N)$.

The one-body density matrix $g_1(z)$ describes the spatial correlations in the wave
function and is defined as
\begin{eqnarray}
g_1(z) = \frac{N}{n}
\frac{\int\Psi_0^*(z_1+z, ..., z_N )\Psi_0(z_1, ..., z_N)\, dz_2... dz_N}
{\int|\Psi_0(z_1, ..., z_N)|^2\, dz_1... dz_N} \;,
\label{g1}
\end{eqnarray}
where $n = N/L$ is the one-dimensional density. The normalization of $g_1(z)$ is
chosen in such a way that $g_1(0) = 1$.

Another important quantity is the pair distribution function $g_2(|z_1-z_2|)$, which
corresponds to the probability of finding two particles separated by $|z_1-z_2|$:
\begin{eqnarray}
g_2(|z_1-z_2|) = \frac{N(N-1)}{n^2}
\frac{\int|\Psi_0(z_1, ..., z_N)|^2\, dz_3... dz_N}{\int|\Psi_0(z_1, ..., z_N)|^2\, dz_1... dz_N} \;.
\label{g2}
\end{eqnarray}

The normalization is chosen in such a way that at large distances $g_2(z)$ goes to
$1-\frac{1}{N}$, {\it i.e.} becomes unity in the thermodynamic limit $N\to\infty$.

The value at zero distance of the three-body correlation function gives the probability of
finding three particles at the same position in space
\begin{eqnarray}
g_3(0) = \frac{N(N-1)(N-2)}{n^3}
\frac{\int|\Psi_0(0, 0, 0, z_4, ..., z_N)|^2\, dz_4... dz_N}
{\int|\Psi_0(z_1, ..., z_N)|^2\, dz_1... dz_N} \;.
\label{g3}
\end{eqnarray}

The knowledge of the density dependence of $g_3(0)$ allows one to estimate the rate of 
three-body recombinations which is of great experimental relevance~\cite{Laburthe Tolra}.

Much useful information can be obtained from the Fourier transform of the above correlation functions.
The momentum distribution $n(k)$ is related to the one-body
density matrix [Eq.~(\ref{g1})]:
\begin{eqnarray}
n(k) = n\int e^{ikz} g_1(z)\,dz \;,
\label{nk}
\end{eqnarray}
and the static structure factor $S(k)$ is instead related to the pair distribution
function [Eq.~(\ref{g2})]:
\begin{eqnarray}
S(k) = 1 + n\int e^{ikz} (g_2(z)-1)\,dz \;.
\label{Fourier Sk}
\end{eqnarray}

The momentum distribution can be measured in time of flight experiments
\cite{Paredes} and the static structure factor using Bragg
spectroscopy \cite{BraggSpectroscopy}.

\section{\label{LL} Lieb-Liniger Hamiltonian}

A bosonic gas at $T = 0$, confined in a waveguide or in a very elongated harmonic
trap, can be described in terms of a one-dimensional model if the energy of the
motion in the longitudinal direction is insufficient to excite the levels of the
transverse confinement. Further, if the range of the interatomic potential is much
smaller than the interparticle distance and the characteristic length of the
external confinement, a single parameter is sufficient to describe interactions,
namely the effective one-dimensional scattering length $a_{1D}$. In this case the
particle-particle interactions can be safely modeled by a $\delta$-pseudopotential.
Such a system is described by the Lieb-Liniger (LL) Hamiltonian \cite{LL1}:
\begin{eqnarray}
\hat H_{LL} = -\frac{\hbar^2}{2m}\sum\limits_{i=1}^N\frac{\partial^2}{\partial z^2_i}
+g_{1D}\sum\limits_{i<j}\delta(z_i-z_j) \;,
\label{HLL}
\end{eqnarray}
where the coupling constant $g_{1D}$ is related to $a_{1D}$ by $g_{1D} =
-2\hbar^2/ma_{1D}$, $m$ being the particle mass. In the presence of a
tight harmonic transverse confinement, characterized by the oscillator length
$a_\perp = \sqrt{\hbar/m\omega_\perp}$, the scattering length $a_{1D}$ was shown to
exhibit a non trivial behavior in terms of the 3D $s$-wave scattering length
$a_{3D}$ due to virtual excitations of transverse oscillator levels \cite{Olshanii g1D}
\begin{eqnarray}
a_{1D} = -a_\perp\left(\frac{a_\perp}{a_{3D}}-1.0326\right) \;.
\label{a1D}
\end{eqnarray}

In typical experimental conditions, far from a Feshbach resonance, one has
$a_{3D}\ll a_\perp$. In this case the above equation simplifies to
$a_{1D} = -a_\perp^2/a_{3D}$, which coincides with the mean-field prediction~\cite{Petrov}. In the
vicinity of a magnetic Feshbach resonance the value of $a_{3D}$ can become
comparable with $a_\perp$ and the coupling constant $g_{1D}$ varies over a wide
range as it goes through a confinement induced resonance \cite{Olshanii g1D}.

All properties in the model depend only on one parameter, the dimensionless density
$n|a_{1D}|$. Contrary to the 3D case, where at low density the gas is weakly
interacting, in 1D small values of the gas parameter $n|a_{1D}|$ correspond to
strongly correlated systems. This peculiarity of 1D systems can be easily understood
by comparing the characteristic kinetic energy $\hbar^2n^{2/D}/2m$, where $D$
denotes the dimensionality, to the mean-field interaction energy $g n$. In 3D,
$g_{3D}n\propto n|a_{3D}|\ll n^{2/3}$ if $n|a_{3D}|^3\ll 1$. In 1D, instead,
$g_{1D}n\propto n/|a_{1D}|\gg n^2$ if $n|a_{1D}|\ll 1$.

\begin{figure}
\includegraphics[angle=-90,width=0.4\columnwidth]{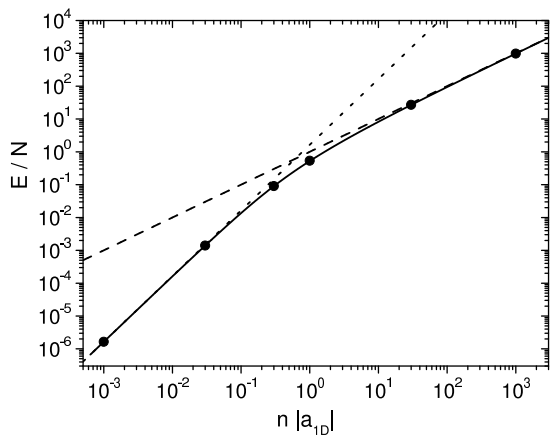}
\caption{Energy per particle: Bethe {\it ansatz} solution (solid line);
GP limit (dashed line); TG limit (dotted line). The circles are the results of DMC calculations.
Energies are in units of $\hbar^2/(ma_{1D}^2)$.}
\label{Fig Energy}
\end{figure}

The ground-state energy of the Hamiltonian (\ref{HLL}) with $g_{1D}>0$ was first
obtained by Lieb and Liniger \cite{LL1} using the Bethe {\it ansatz} method. The
energy of the system is conveniently expressed as $E/N = e(n|a_{1D}|)\hbar^2n^2/2m$,
where the function $e(n|a_{1D}|)$ is obtained by solving a system of integral
equations. In the mean-field Gross-Pitaevskii (GP) regime, $n|a_{1D}|\gg 1$, the
energy per particle is linear in the density $E^{GP}/N = g_{1D} n/2$, while in the
strongly correlated Tonks-Girardeau (TG) regime, $n|a_{1D}|\ll 1$, the dependence is quadratic $E^{TG}/N
= \pi^2\hbar^2n^2/6m$. The equation of state resulting from a numerical solution of
the LL integral equations is shown in Fig.~\ref{Fig Energy} as a function of the gas
parameter $n|a_{1D}|$.

In the TG regime, the energy of an incident particle is not
sufficient to tunnel through the repulsive interaction potential and two particles will
never be at the same position in space. This constraint, together with the spatial peculiarity
of 1D systems, acts as an effective Fermi exclusion principle. Indeed, in this
limit, the system of bosons acquires many Fermi-like properties. There exists a
direct mapping of the wave function of strongly interacting bosons onto a wave
function of non interacting spinless fermions due to Girardeau \cite{Girardeau}. 
The chemical potential, $\mu=\partial E/\partial N$, of a TG gas equals the 1D Fermi energy 
$\mu=\hbar^2k_F^2/2m$, where $k_F=\pi n$ is the Fermi wave vector, and the
speed of sound $c$, defined through the inverse compressibility
$mc^2=n\partial\mu/\partial n$, is given by $c =\hbar k_F/m$.

Further, the pair distribution function
\begin{eqnarray}
g_2(z) = 1 - \frac{\sin^2 \pi n z}{(\pi n z)^2} \;,
\label{g2TG}
\end{eqnarray}
which exhibits Friedel-like oscillations, and the static structure factor of the TG
gas
\begin{equation}
S(k)=\left\{
\begin{array}{cc}
|k|/(2\pi n),&|k|<2\pi n\\
1,&|k|>2\pi n \;,\\
\end{array}
\right.
\label{SkTG}
\end{equation}
can be calculated exactly exploiting the Bose-Fermi mapping~\cite{Girardeau}.

The one-body density matrix $g_1(z)$ of a TG gas has been calculated in terms of
series expansions holding at small and large distances in Ref.~\cite{Corr tonks}. The
leading long-range term decays as
\begin{eqnarray}
g_1(z) = \frac{\sqrt{\pi e}2^{-1/3}A^{-6}}{\sqrt{|z|n}} \;,
\label{g1TG}
\end{eqnarray}
($A=1.28...$ is Glaisher's constant) and yields an infrared divergence in the
momentum distribution $n(k) \propto 1/\sqrt{|k|/n}$.

Outside the TG regime full expressions of the correlation functions are not known.
The long-range asymptotics can be calculated using the hydrodynamic theory of the
low-energy phonon-like excitations \cite{Reatto, Schwartz, Korepin}. For $g_1(z)$ one
finds the power-law decay
\begin{eqnarray}
g_1(z) = \frac{C_{asympt}}{|z n|^\alpha} \;,
\label{g1LR}
\end{eqnarray}
where $\alpha = mc/(2\pi\hbar n)$ and $C_{asympt}$ is a numerical coefficient. This
result holds for distances $|z|\gg\xi$, where $\xi=\hbar/(\sqrt{2}mc)$ is the
healing length. In the TG regime $c = \pi\hbar n/m$ and $\alpha=1/2$ as anticipated
above. In the opposite GP regime ($n|a_{1D}|\gg 1$) one finds $\alpha =
1/(\pi\sqrt{2n|a_{1D}|})$, yielding a vanishingly small value for $\alpha$. The
power-law decay of the one-body density matrix excludes the existence of
Bose-Einstein condensation in infinite systems \cite{Schultz}.
The behavior of the momentum distribution for $|k|\ll 1/\xi$ follows immediately
from Eq.~(\ref{g1LR}):
\begin{eqnarray}
n(k) = C_{asympt} \left|\frac{2n}{k}\right|^{1-\alpha}
\frac{\sqrt\pi\Gamma\left(\frac{1}{2}-\frac{\alpha}{2}\right)}{\Gamma\left(\frac{\alpha}{2}\right)} \;,
\label{nk small}
\end{eqnarray}
where $\Gamma(z)$ is the Gamma function.

The hydrodynamic theory allows one to calculate also the static structure factor in the long-wavelength 
regime, $|k|\ll 1/\xi$. One finds the well-known Feynmann
result \cite{Feynmann}
\begin{eqnarray}
S(k) = \frac{\hbar|k|}{2mc} \;.
\label{SkSR}
\end{eqnarray}

Recently, the short-range behavior of the one-, two-, and three-body correlation
functions has also been investigated. The value at $z=0$ of the pair correlation
function for arbitrary densities can be obtained from the equation of state through
the Hellmann-Feynman theorem \cite{Gangardt}:
\begin{eqnarray}
g_2(0)=-\frac{(n|a_{1D}|)^2}{2} e' \;, 
\label{g2SR}
\end{eqnarray}
where the derivative of the equation of state $e(n|a_{1D}|)$ is to be taken with respect to $n|a_{1D}|$.

The $z=0$ value of the three-body correlation function was obtained within a
perturbation scheme in the regions of strong and weak interactions \cite{Gangardt}.
It is very small in the TG limit ($n|a_{1D}|\ll 1$)
\begin{eqnarray}
g_3(0) = \frac{(\pi n |a_{1D}|)^6}{60} \;,
\label{g3TG}
\end{eqnarray}
and goes to unity in the GP regime ($n|a_{1D}|\gg 1$)
\begin{eqnarray}
g_3(0) = 1-\frac{6\sqrt{2}}{\pi\sqrt{n|a_{1D}|}} \;.
\label{g3GP}
\end{eqnarray}

The first terms of the short-range expansion of $g_1(z)$ can also be calculated from
the knowledge of the equation of state \cite{Olshanii}
\begin{eqnarray}
g_1(z) = 1 - \frac{1}{2}(e+e' n|a_{1D}|)(nz)^2 + \frac{e'}{6}(n|z|^3) \;,
\label{g1SR}
\end{eqnarray}
holding for arbitrary densities and for small distances $n|z| \ll 1$.

\section{Quantum Monte Carlo Method \label{SMC}}

We use Variational Monte Carlo (VMC) and Diffusion Monte Carlo (DMC) methods in
order to study the ground-state properties of the system. In VMC one calculates the
expectation value of the Hamiltonian over a trial wave function $\psi_T({\bf R}, A,
B, ...)$, where ${\bf R}=(z_1,...,z_N)$ denotes the particle coordinates and $A$, $B$,
... are variational parameters. According to the variational principle, the energy
\begin{eqnarray}
E_{VMC} = \frac{\langle\psi_T|\hat H|\psi_T\rangle}{\langle\psi_T|\psi_T\rangle}
\label{Evmc}
\end{eqnarray}
provides an upper bound to the ground-state energy, $E_{VMC}\ge E_0$. The
variational parameters $A,B,...$ are optimized by minimization of the variational
energy (\ref{Evmc}).

In order to remove the bias in the estimate of the ground-state energy caused by the
particular choice of the trial wave function, we resort to the DMC method, which allows
one to solve exactly, apart from statistical uncertainty, the many-body
Schr\"odinger equation of a Bose system at zero temperature~\cite{BORO1}. The evolution in
imaginary time, $\tau = i t/\hbar$, is performed for the product $f({\bf
R},\tau)=\psi_T({\bf R})\Psi({\bf R},\tau)$, where $\Psi({\bf R},\tau)$ denotes the
wave function of the system and $\psi_T({\bf R})$ is a trial function used for
importance sampling. The time-dependent Schr\"odinger equation for the function
$f({\bf R},\tau)$ can be written as
\begin{eqnarray}
-\frac{\partial f({\bf R},\tau)}{\partial\tau}= &-& D\nabla_{\bf R}^2 f({\bf R},\tau) + D \nabla_{\bf R}[{\bf F}({\bf R})
f({\bf R},\tau)]
+ [E_L({\bf R})-E_{ref}]f({\bf R},\tau) \;,
\label{DMCalgorithm}
\end{eqnarray}
where $E_L({\bf R}) = \psi_T({\bf R})^{-1}H\psi_T({\bf R})$ denotes the local
energy, ${\bf F}({\bf R})=2\psi_T({\bf R})^{-1}\nabla_{\bf R} \psi_T({\bf R})$ is
the quantum drift force, $D=\hbar^2/(2m)$ plays the role of an effective diffusion
constant, and $E_{ref}$ is a reference energy introduced to stabilize the numerical
evaluation. The energy and other observables of the state of the system are
calculated from averages over the asymtpotic distribution function $f({\bf
R},\tau\to\infty)$. It is easy to check by decomposing $f({\bf R},\tau)$ on the
basis of stationary states of the Hamiltonian, that contributions of the excited
states vanish exponentially fast with the imaginary time $\tau$ and asymptotically
one obtains $\lim_{\tau\to\infty}f({\bf R},\tau) = \psi_T({\bf R})\Psi_0({\bf R})$
for all trial wave functions non orthogonal to the ground-state wave function
$\Psi_0({\bf R})$. The local energy $E_L$ sampled over the asymptotic distribution
equals exactly the ground-state energy
\begin{eqnarray}
\lim_{\tau\to\infty}
\frac{\int E_L({\bf R})f({\bf R},\tau)\,d{\bf R}}{\int f({\bf R},\tau)\,d{\bf R}}
= \frac{\langle\psi_T|\hat H|\Psi_0\rangle}{\langle\psi_T|\Psi_0\rangle} = E_0 \;.
\label{Edmc}
\end{eqnarray}

In the present study the trial wave function is chosen of the the Bijl-Jastrow form:
\begin{eqnarray}
\psi_T(z_1, ..., z_N) = \prod_{i<j}f(z_{ij}),
\end{eqnarray}
where $f(z)$ is a two-body term chosen as
\begin{equation}
f(z)=\left\{
\begin{array}{lr}
A \cos [k(|z|-B)], &|z|<Z\\
\sin^\beta (\pi|z|/L),&|z|>Z \;.\\
\end{array}
\right.
\label{ftrial}
\end{equation}
We consider $N$ particles in a box of size $L$ with periodic boundary conditions. In
the construction of the trial wave function we have ensured that $f(z)$ is uncorrelated 
at the box boundaries, $f(z=\pm L/2) = 1$ and the derivative $f'(z=\pm L/2) = 0$. 
For $|z|<Z$, the Bijl-Jastrow term $f(z)$ corresponds to the exact solution of the two-body 
problem with
the potential $g_{1D}\delta(z)$ and provides a correct description of short-range
correlations. Long-range correlations arising from phonon excitations are instead
accounted for by the functional dependence of $f(z)$ for $|z|>Z$ \cite{Reatto}. The
$z=0$ boundary condition $f'(0^+) - f'(0^-) = 2 f(0)/|a_{1D}|$, which accounts for
the $\delta$-function potential, fixes the parameter $k$ through the relation
$k|a_{1D}|\tan kB = 1$. The remaining parameters $A,B$ and $\beta$ are fixed by the
continuity conditions at the matching point $z = Z$ of the function $f(z)$, its
derivative $f'(z)$ and the local energy $-2f''(z)/f(z)$. The value of the matching
point $Z$ is a variational parameter which we optimize using VMC. The TG wave
function of Ref.~\cite{Girardeau}, $\Psi_0^{TG} = \prod_{i<j} |\sin[\pi(z_i-z_j)/L]|$, is
obtained as a special case of our trial wave function $\psi_T$ for $Z = B = L/2$ and
$kL = \pi$.

The choice of a good trial wave-function is crucial for the efficiency of the
calculation. In order to prove that our trial wave-function is indeed very close to
the true ground-state $\Psi_0({\bf R})$, we compare in Table~I the variational
energy $E_{VMC}$ with the exact solution based on the use of the Bethe {\it ansatz}
\cite{LL1}. The corresponding results obtained using DMC coincide, within
statistical uncertainty, with the exact ones and are shown in Fig.~\ref{Fig Energy}.

\begin{table}
\centering
\begin{tabular}{|l|l|l|}
\hline
$n|a_{1D}|$&$E_{LL}/N$&$E_{VMC}/N$\\
\hline
$10^{-3} $&$1.6408\cdot 10^{-6} $&$1.6415(1)\cdot 10^{-6}$\\
$0.03    $&$1.3949\cdot 10^{-3} $&$1.3957(3)\cdot 10^{-3}$\\
$0.3     $&$9.0595\cdot 10^{-2} $&$9.093(1)\cdot 10^{-2}$\\
$1       $&$0.5252              $&$0.5259(1)     $\\
$30      $&$26.842              $&$27.19(5)    $\\
$10^3    $&$981.15              $&$983.6(3)    $\\
\hline
\end{tabular}
\label{table1}
\caption{Energy per particle for different values of the gas parameter $n|a_{1D}|$:
$E_{LL}$ --- exact result obtained by solving the Lieb-Liniger equations, $E_{VMC}$
--- variational result, Eq. (\ref{Evmc}), obtained from optimization of the trial wave
function (\ref{ftrial}).}
\end{table}

Besides the ground-state energy $E_0$, the DMC method gives exact results also for
local correlation functions, such as the pair distribution function $g_2(z)$ and the
three-body correlator $g_3(0)$, for which one can use the method of ``pure''
estimators \cite{BORO2}. Instead, in the calculation of the non-local one-body
density matrix $g_1(z)$, the bias from the trial wave function can be reduced using
the extrapolation technique: $\langle\Psi_0|\hat A|\Psi_0\rangle=
2\,\langle\Psi_0|\hat A|\psi_T\rangle-\langle\psi_T|\hat A|\psi_T\rangle$, written
here for a generic operator $\hat A$. The ``mixed'' estimator $\langle\Psi_0|\hat
A|\psi_T\rangle$ is the direct output of the DMC calculation and the variational
estimator $\langle\psi_T|\hat A|\psi_T\rangle$ is obtained from a VMC calculation.
This procedure is accurate only if $\psi_T \simeq\Psi_0$. In the present study DMC
and VMC give results for $g_1(z)$ which are very close and we believe that the
extrapolation technique removes completely the bias from $\Psi_T({\bf R})$.

Calculations are carried out for a finite number of particles $N$. In order to
extrapolate to infinite systems, we increase $N$ and study the
convergence in the quantities of interest. The dependence on the number of particles
(so called finite-size effect) is more pronounced in the regime $n|a_{1D}|\gg 1$,
where correlations extend to very large distances. Finite size effects can be best
investigated by considering the one-body density matrix. In Fig.~\ref{Fig
finitesize} we show $g_1(z)$ at a fixed density, $n|a_{1D}| = 30$, for systems of $N
= 50, 100, 200, 500$ particles and we compare the long-range behavior with the
power-law decay $g_1(z)\propto 1/(n|z|)^\alpha$. For all values of $N$ deviations
from a power-law decay are visible close to the boundary of the box, $z = L/2$, due to
the use of periodic boundary conditions. We find that the slope of $g_1(z)$ at large
distances depends on $N$, approaching the predicted hydrodynamic value as $N$ increases. 
For $N =500$ we recover the result $\alpha = 0.04$. For smaller values of $n|a_{1D}|$,
finite-size effects are less visible and, for practical purposes, calculations with
$N < 500$ are sufficient.
\begin{figure}
\includegraphics[angle=-90,width=0.4\columnwidth]{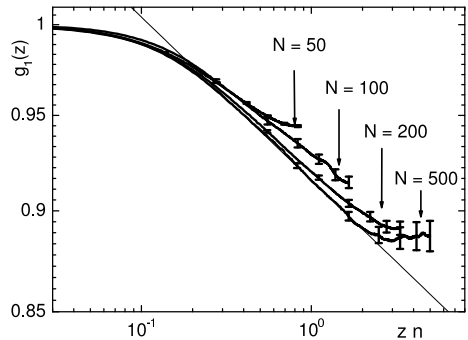}
\caption{Study of finite-size effects in the calculation of the one-body
density matrix $g_1(z)$ at $n|a_{1D}| = 30$. The thin solid line corresponds to the
power-law decay $A/(n|z|^{0.04})$ with $A$ obtained from a best fit to the $N=500$
result. }
\label{Fig finitesize}
\end{figure}

\section{Results \label{results}}

\begin{figure}
\includegraphics[angle=-90,width=0.4\columnwidth]{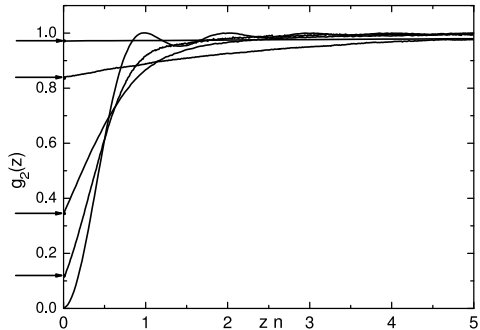}
\caption{Pair distribution function for different values of the gas
parameter. In ascending order of the value at zero $n|a_{1D}| = 10^{-3}, 0.3, 1, 30,
10^3$. Arrows indicate the value of $g_2(0)$ as obtained from Eq.~(\ref{g2SR}).}
\label{Fig g2}
\end{figure}

We calculate the pair distribution function $g_2(z)$ for densities ranging from very
small values of the gas parameter $n|a_{1D}|\ll 1$ (TG regime) up to $n|a_{1D}|\gg
1$ (GP regime). The results are presented in Fig.~\ref{Fig g2}. In the GP regime the
correlations between particles are weak and $g_2(z)$ is always close to the
asymptotic value $g_2(|z|\to\infty)=1$. By decreasing $n|a_{1D}|$ (thus making the
coupling constant $g_{1D}$ larger) we enhance beyond-mean field effects and the role
of correlations. For the smallest considered value of the gas parameter $n|a_{1D}| =
10^{-3}$, we see oscillations in the pair distribution function, which is a
signature of strong correlations present in the gas. At the same density we compare
the pair distribution function with the one corresponding to a TG gas,
Eq.~(\ref{g2TG}), finding no visible difference.

In the same figure we show the analytical predictions for the value of $g_2(0)$,
Eq.~(\ref{g2SR}). In the TG regime particles are never at the same position and
consequently $g_2(0) = 0$. With a weaker interaction between particles, we find a
finite probability that two particles come close to each other in agreement with
Eq.~(\ref{g2SR}). As we go further towards the GP regime, the interaction potential
becomes more and more transparent and we approach the ideal gas limit $g_2(0) = 1$.

\begin{figure}
\includegraphics[angle=-90,width=0.4\columnwidth]{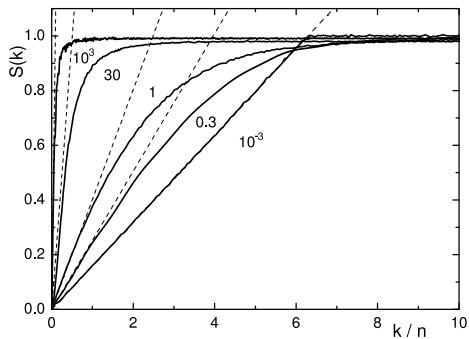}
\caption{Static structure factor at the density $n|a_{1D}| = 10^{-3}, 0.3, 1, 30,
10^3$ (solid lines). The dashed lines are the corresponding long-wavelength
asymptotics from Eq.~(\ref{SkSR}).}
\label{Fig Sk}
\end{figure}

In Fig.~\ref{Fig Sk} we present results of the static structure factor obtained from
$g_2(z)$ according to Eq.~(\ref{Fourier Sk}). At the smallest density, $n|a_{1D}| =
10^{-3}$, our results are indistinguishable from the $S(k)$ of a TG gas
[Eq.~(\ref{SkTG})]. For all densities the small wave vector part of $S(k)$ is
dominated by phononic excitations. We compare the DMC results with the asymptotic
linear slope [Eq.~(\ref{SkSR})]. We see that in the strongly correlated regime the
phononic contribution provides a correct description of $S(k)$ up to values of $k$
of the order of the inverse mean interparticle distance $n$. In the GP regime the healing
length becomes significantly larger than the mean interparticle distance, leading to
deviations from the linear slope for smaller values of $k$.

\begin{figure}
\includegraphics[angle=-90,width=0.4\columnwidth]{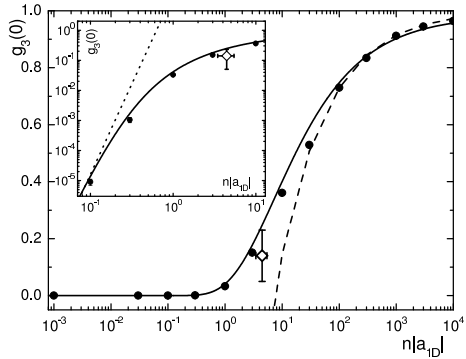}
\caption{Value at zero distance of the three-body correlation function $g_3(0)$ (circles);
GP limit, Eq.~(\ref{g3GP}), dashed line; mean-field factorization, $g_3(0)=(g_2(0))^3$
solid line. Inset: small density region on a log-log scale: TG limit,
Eq.~(\ref{g3TG}), dotted line. Open symbol: measured value of $g_3(0)$
from three-body loss experiments \cite{Laburthe Tolra}} 
\label{Fig g3}
\end{figure}

We show in Fig.~\ref{Fig g3} the results for the value at zero distance of the three-body
correlation function, Eq.~(\ref{g3}), calculated over a large range of densities. At
large density, $n|a_{1D}| = 10^4$, the probability of three-body collisions is large
and the result of Bogoliubov theory, Eq.~(\ref{g3GP}), provides a good description
of $g_3(0)$. By reducing $n|a_{1D}|$ the value of $g_3(0)$ decreases, becoming
vanishingly small for values of the gas parameter $n|a_{1D}|\ll 1$. In order to
resolve the dependence of $g_3(0)$ on the density in the TG regime, we plot the
results on a log-log scale (inset of Fig.~\ref{Fig g3}). We obtain that $g_3(0)$ is
proportional to the fourth power of the gas parameter in agreement with
Eq.~(\ref{g3TG}). A reliable evaluation of the three-body correlator for small
densities is difficult due to the very small value of $g_3(0)$ itself. It is
interesting to notice that $g_2^3(0)$ is close to $g_3(0)$ over the whole density
range. This estimate of $g_3(0)$ has been discussed in Ref.~\cite{Laburthe Tolra}. The
coefficient of three-body losses has been measured in quasi-1D
configurations realized with deep two-dimensional optical lattices \cite{Laburthe
Tolra}. The value of $g_3(0)$ extracted from these measurements is also shown in
Fig.~\ref{Fig g3}.

\begin{figure}
\includegraphics[angle=-90,width=0.4\columnwidth]{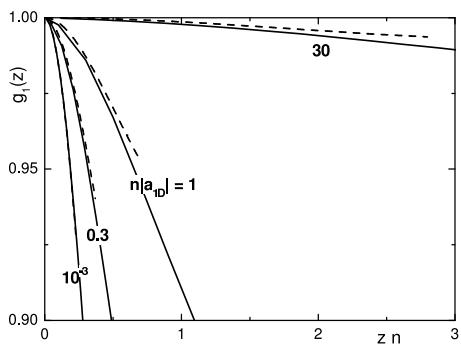}
\caption{Short range behavior of the one-body density matrix
at density $n|a_{1D}| = 10^{-3}, 0.3, 1, 30$ (solid lines), compared with the series
expansion at small distances [Eq.~(\ref{g1SR})] (dashed lines).}
\label{Fig g1small}
\end{figure}

\begin{figure}
\includegraphics[angle=-90,width=0.4\columnwidth]{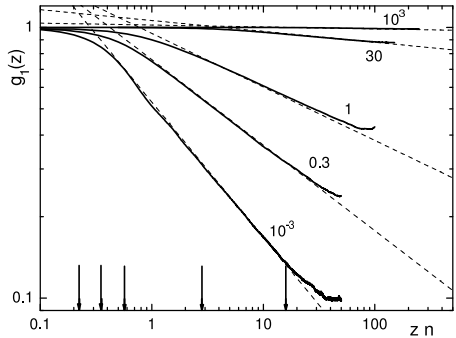}
\caption{Long-range behavior of the one-body density matrix
(solid lines), best fits to the long-wavelength asymptotics from Eq.~(\ref{g1LR})
(dashed lines). Values of the density are $n|a_{1D}| = 10^{-3}, 0.3, 1, 30, 10^{3}$.
The arrows indicate the value of the product of the density and the healing length
$\xi n$: the leftmost corresponds to $n|a_{1D}| = 10^{-3}$, the rightmost to
$n|a_{1D}| = 10^3$}
\label{Fig g1}
\end{figure}

We calculate the spatial dependence of the one-body density matrix, Eq.~\ref{g1},
for different values of the gas parameter. At small distances we compare the DMC results with the
short-range expansion, Eq.~(\ref{g1SR}), finding agreement for $nz\ll 1$ (see
Fig.~\ref{Fig g1small}). For distances larger than the healing length $\xi$ we expect the
hydrodynamic theory to provide a correct description. The long-range decay shown in
Fig.~\ref{Fig g1} exhibits a power-law behavior in agreement with the prediction of
Eq.~(\ref{g1LR}). The coefficient of proportionality in Eq.~(\ref{g1LR}) is fixed by
a best fit to the DMC results. The small deviations from the power law-decay at the
largest distances ($z\approx L/2$) are due to finite size effects (see Fig.~\ref{Fig
finitesize}).

In the weakly interacting GP regime the coefficient $C_{asympt}$ of Eq. (\ref{g1LR}) can be calculated
from a hydrodynamic approach \cite{Casympt}. One obtains
\begin{eqnarray}
C_{asympt} = \left(\frac{e^{1-\gamma}}{8\pi\alpha}\right)^\alpha(1+\alpha) \;,
\label{g1MF}
\end{eqnarray}
where $\gamma=0.577$ is Euler's constant and $\alpha=mc/(2\pi\hbar n)$.

Although result (\ref{g1MF}) is formally derived in the weakly interacting limit,
$\alpha\ll 1$, it works well in the whole range of densities. Indeed, the value of
$C_{asympt}$ obtained from the best fit to $g_{1}(z)$, as shown in Fig.~\ref{Fig
g1}, is always in good agreement with the prediction (\ref{g1MF}). For example, in
the strongly-interacting TG regime the comparison between Eq. (\ref{g1MF}) and the exact
result in Eq. (\ref{g1TG}) gives only $0.3 \%$ difference. A different expression was
obtained by Popov \cite{Popov} (and later recovered in \cite{Castin}) giving
$C^{Popov}_{asympt} = \left(\frac{e^{2-\gamma}}{8\pi\alpha}\right)^\alpha$. Both
expressions coincide for small values of $\alpha$, but Popov's coefficient yields
larger errors approaching the TG regime, with $10\%$ maximal error, as it was pointed
out in Ref.~\cite{Cazalilla}. A comparison of the different coefficients is
presented in Table~II.

\begin{table}
\centering
\begin{tabular}{|l|l|l|l|}
\hline
$n|a_{1D}|$ & $C^{DMC}_{asympt}$ &  $C^{Popov}_{asympt}$ & $C_{asympt}$\\
\hline
1000  & 1.02  & 1.0226 & 1.0226 \\
30    & 1.06  & 1.0588 & 1.0579 \\
1     & 0.951 & 0.9646 & 0.9480 \\
0.3   & 0.760 & 0.8145 & 0.7814 \\
0.001 & 0.530 & 0.5746 & 0.5227 \\
\hline
\end{tabular}
\label{table2}
\caption{Coefficient of the long-range decay of the one-body density matrix
as defined in Eq.~(\ref{g1LR}). The first column is the one-dimensional gas
parameter, the second column is the coefficient extracted from the best fit to the DMC
results (see Fig.~\ref{Fig g1}), the third column is Popov's prediction, and the
fourth column is Eq.~(\ref{g1MF}). The value of the gas parameter $n|a_{1D}| =
0.001$ corresponds to the deep TG regime where one can apply Eq.~(\ref{g1TG})
yielding $C^{TG}_{asympt} = 0.5214$.}
\end{table}

\begin{figure}
\includegraphics[angle=-90,width=0.4\columnwidth]{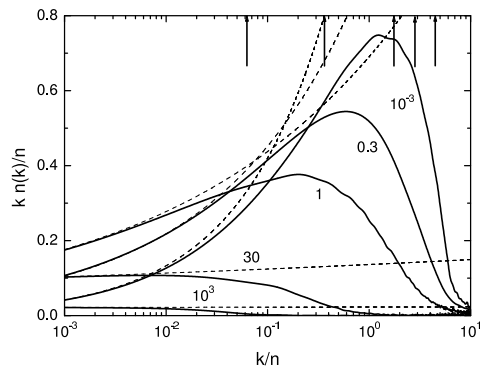}
\caption{Momentum distribution at density $n|a_{1D}| = 10^{-3}, 0.3, 1, 30, 10^{3}$.
The dashed lines correspond to the infrared behavior of Eq.~(\ref{nk small}). The
arrows indicate the value of $1/\xi n$: the rightmost corresponds to $n|a_{1D}| =
10^{-3}$, the leftmost to $n|a_{1D}| = 10^3$.}
\label{Fig nk}
\end{figure}

The momentum distribution, Eq.~(\ref{nk}), is obtained from the Fourier
transform of the calculated one-body density matrix at short distances and the best fitted
power-law decay at large distances. The momentum distribution exhibits the infrared
divergence of Eq.~(\ref{nk small}). We present the results for $n(k)$ by plotting in
Fig.~\ref{Fig nk} the combination $k n(k)$, where the divergence is absent. We
notice that the infrared asymptotic behavior is recovered for values of $k$
considerably smaller than the inverse healing length $1/\xi$. At large $k$ the
numerical noise of our results is too large to extract evidences of the $1/k^4$
behavior predicted in Ref.~\cite{Olshanii}.

\section{Conclusions\label{SC}}

This paper presents a thorough study of the correlation functions in a
one-dimensional homogeneous Bose gas described by the Lieb-Liniger Hamiltonian. The
correlation functions are calculated for all interaction regimes using exact quantum Monte Carlo
methods. The results on the pair distribution function, one-body density matrix and
their Fourier transformations have already been presented in Ref.~\cite{us}
and are briefly reviewed here. We carry out a more detailed study of the short- and
long-range behavior of the one-body density matrix, including the comparison with
analytic expansions. We also calculate the probability of finding three particles at
the same spatial position as a function of the interaction strength and we compare
it with the asymptotic results holding in the weakly- and strongly-interacting
regime and with available experimental results from three-body loss measurements.

This research is supported by the Ministero dell'Istruzione, dell'Universit\`a e
della Ricerca (MIUR).

\end{document}